\documentclass[aps,pre,superscriptaddress,twocolumn,amsmath,amssymb,showpacs]{revtex4}

\usepackage[T1]{fontenc}
\usepackage[latin1]{inputenc}

\usepackage{graphicx}

\usepackage{dcolumn}

\usepackage{upgreek}

\makeatletter

\begin{document}

\title{Entropic Stochastic Resonance}

\author{P.S. Burada}
\affiliation{Institut f\"ur Physik,
  Universit\"at Augsburg,
  Universit\"atsstr. 1,
  D-86135 Augsburg, Germany}

\author{G. Schmid}
\affiliation{Institut f\"ur Physik,
  Universit\"at Augsburg,
  Universit\"atsstr. 1,
  D-86135 Augsburg, Germany}

\author{D. Reguera}
\affiliation{Departament de F\'isica Fonamental,
  Facultat de F\'isica,
  Universidad de Barcelona,
  Diagonal 647, E-08028 Barcelona, Spain}

\author{M.H. Vainstein}
\affiliation{Departament de F\'isica Fonamental,
  Facultat de F\'isica,
  Universidad de Barcelona,
  Diagonal 647, E-08028 Barcelona, Spain}

\author{J.M. Rubi}
\affiliation{Departament de F\'isica Fonamental,
  Facultat de F\'isica,
  Universidad de Barcelona,
  Diagonal 647, E-08028 Barcelona, Spain}

\author{P. H\"anggi}
\affiliation{Institut f\"ur Physik,
  Universit\"at Augsburg,
  Universit\"atsstr. 1,
  D-86135 Augsburg, Germany}

\date{\today}


\begin{abstract}
We present a novel scheme for the appearance of Stochastic Resonance
when the dynamics of a Brownian particle takes place in a confined medium.
The presence of uneven boundaries, giving rise to an entropic contribution
to the potential, may upon application of a periodic driving force result
in an increase of the spectral amplification at an optimum value of
the ambient noise 
level. This {\it Entropic Stochastic Resonance} (ESR), characteristic
of small-scale systems,
may constitute a useful mechanism for the manipulation and control of
single-molecules and nano-devices.
\end{abstract}


\maketitle


{\itshape Stochastic Resonance} (SR) describes the counterintuitive
phenomenon where an appropriate dose of  noise is not  harmful for
the detection or transduction of an incoming, generally weak signal,
but rather of constructive use in the sense that a weak signal becomes amplified upon harvesting the ambient
noise in metastable, nonlinear systems \cite{gammaitoni}.
Since its first discovery in the early
eighties SR has been observed in a great variety of
systems pertaining to different disciplines such as physics, chemistry,
engineering, biology and biomedical
sciences
\cite{PT_SR,gammaitoni,chemphyschem,vilar_mono,Lutz99,Schmid01,BuchSR,Yasuda08,scholarpediaSR,GoychukSR}.
The list of models and applications is still growing.
In particular, SR has found widespread interests and 
applications within biological physics.

The research on SR has primarily been focused on systems with purely energetic
potentials. However, in situations frequently found in soft condensed
matter and biological systems, particles move in constrained regions such
as small cavities, pores or channels whose presence and shape play an important role
for the SR-dynamics \cite{GoychukSR}, sometimes even more important than the well-studied case
of energetic barriers in such systems \cite{hille,zeolites,liu,berzhkovski}.
In this work, we demonstrate that irregularities in
the form of confining, curved  boundaries, being modeled via an entropic potential, can cause
noise-assisted, resonant-like behaviors in the system under consideration. Confinement, an inherent property of
small-scale systems, can thus constitute an important source of noise-induced resonant
effects with interesting applications in the design and control of these
systems.

\begin{figure}[t]
  \centering
  \includegraphics{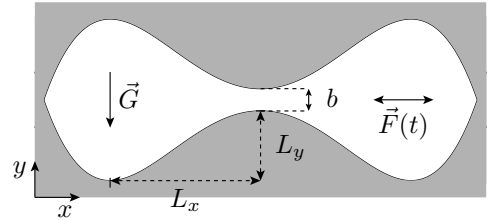}
  \caption{Schematic illustration of the two-dimensional
    structure confining the motion of the Brownian particles.
    The symmetric structure is defined by a quartic double well function,
    cf. Eq.~\eqref{eq:widthfunctions}, involving the geometrical
    parameters $L_{x}$, $L_{y}$ and $b$.
    Brownian particles are driven by a sinusoidal
    force $\vec{F}(t)$ along the longitudinal direction and a constant
    force $\vec{G}$ in the transversal direction.
  }
  \label{fig:well}
\end{figure}

The phenomenon of SR is rooted
on a stochastic synchronization between noise-induced hopping events
and the rhythm of the externally applied signal, that taken alone is not
sufficient for the system to overcome a potential barrier. In the
first place, noise enables system transitions and it is in fact responsible
for the observed signal amplification and the emergence of certain degree of order.
In the earliest and basic manifestation of SR, the synchronization of the random
switches of a Brownian particle with a periodic driving force were observed
for a bistable potential.
Moreover, potentials of this type are not only found in systems with
energy barriers, as they may also arise due to the
influence of entropic constraints. Particles diffusing freely in a
confined medium such as the one depicted in Fig.~\ref{fig:well} may give rise to
an activation regime when a constant force $\vec{G}$ in the transverse
direction is imposed.
We will show that the combination of forcing and the presence of entropic
effects deriving from the confinement and the
irregularity of the boundaries give rise to an effective bistable
potential that exhibits
the signatures of Stochastic Resonance.

The dynamics of a particle in a constrained geometry subjected to a sinusoidal
oscillating force $F(t)$ along the axis of the structure and to a constant force $G$ in
the transversal direction can be described by means of the Langevin equation, written
in the overdamped limit as
\begin{align}
  \label{eq:langevin}
  \gamma \, \frac{\mathrm{d}\vec{r}}{\mathrm{d} t} = - G\vec{e_y}
  - F(t)\vec{e_x} +\sqrt{\gamma \, k_{\mathrm{B}}T}\, \vec{\xi}(t)\, ,
\end{align}
where $\vec{r}$ denotes the position of the particle, $\gamma$ is the
friction coefficient, $\vec{e_x}$ and
$\vec{e_y}$ the unit vectors along the $x$ and $y$-directions, respectively, and
$\vec{\xi}(t)$ is a Gaussian white noise with zero mean which
obeys the fluctuation-dissipation relation
$\langle \xi_{i}(t)\,\xi_{j}(t') \rangle = 2\, \delta_{ij}\,
\delta(t - t')$ for $i,j = x,y$.
The explicit form of the longitudinal force is given by
$F(t) = F_0 \sin( \Omega t )$ where $F_0$ is the amplitude and $\Omega$
is the frequency of the sinusoidal driving.

In the presence of confining boundaries, this equation has to be solved by
imposing reflecting boundary conditions
at the walls of the structure. For the 2D structure depicted in
Fig.~\ref{fig:well}, the walls are defined by
\begin{align}
        \label{eq:widthfunctions}
    w_{\mathrm{l}}(x) &= L_y \left(\frac{x}{L_x}\right)^4 -
   2\,L_y\left( \frac{x}{L_x}\right)^2 - \frac{b}{2} \, = -  w_{\mathrm{u}}(x)\, ,
\end{align}
where $ w_{\mathrm{l}}$ and $ w_{\mathrm{u}}$ correspond to the lower
and the upper boundary functions, respectively,
$L_{x}$ corresponds to the
distance between bottleneck position and position of maximal width,
and $L_{y}$ refers to the narrowing of the boundary functions and
$b$ to the remaining width at the bottleneck,
cf. Fig.~\ref{fig:well}.
Consequently, $2\,w(x) = w_{\mathrm {u}}(x) - w_{\mathrm{l}}(x)$
gives the local width of the structure. This choice of the
geometry is intended to resemble the
archetype setup for SR in the context of a  double well potential.
In fact, in the limit of a sufficiently large transversal force, the particle is in
practice restricted to explore the region very close to the lower boundary
of the structure, recovering the effect of an energetic bistable potential.
For sake of a dimensionless description, we henceforth scale all lengths by the
characteristic length $L_x$, i.e. ${\tilde{x}} = x/L_x$, $\tilde{y} =
y/L_{x}$ which implies ${\tilde {b}} = b/L_x$ and
${\tilde {w}}_{\mathrm{l}}= w_{\mathrm{l}}/L_x = - {\tilde {w}}_{\mathrm{u}}$,
time by $\tau=\gamma {L_x}^2/k_{\mathrm{B}}T_{\mathrm{R}}$, the corresponding
characteristic diffusion time at an arbitrary, but irrelevant reference
temperature $T_{\mathrm{R}}$, i.e. ${\tilde {t}} = t/\tau$ and
$\tilde{\Omega} = \Omega \tau$, and
force by $F_{\mathrm{R}}=\gamma L_x/\tau $, 
transversal force ${\tilde {G}} = G/F_{\mathrm{R}}$
and a longitudinally acting, sinusoidal force ${\tilde {F}(\tilde {t})} =
F(t)/F_{\mathrm{R}}$.
In the following we shall omit
the tilde symbols for better legibility.
In dimensionless form the Langevin-equation~\eqref{eq:langevin} and
the boundary functions \eqref{eq:widthfunctions} read:
\begin{align}
  \label{eq:dllangevin}
  \frac{\mathrm{d}\vec{r}}{\mathrm{d} t} & = - G\vec{e_y}
  - F(t)\vec{e_x} +\sqrt{D}\, \vec{\xi}(t)\, ,\\
  \label{eq:dlboundaryfunctions}
  w_{\mathrm{l}}(x) & = -w_{\mathrm{u}}(x) = - w(x)
  = \epsilon x^4 - 2\epsilon x^2
  - b/2 \, ,
\end{align}
where we defined the aspect ratio $\epsilon = L_y/L_x$ and the
dimensionless temperature $D=T/T_{\mathrm{R}}$.

In the absence of a time-dependent applied bias, i.e. $F(t) = 0 $,
it has been shown by a coarsening of the description
\cite{Reguera_PRE,Reguera_PRL,Burada_PRE}
that the Langevin equation \eqref{eq:langevin}
can be reduced to an effective 1D-Fokker-Planck equation, reading in dimensionless form
\begin{align}
\label{eq:fj}
\frac{\partial P(x,t)}{\partial t} =
     \frac{\partial}{\partial x}\left\{D \frac{\partial P}{\partial x} \, +
     V^{\prime}(x,D)\, P \right\} \, ,
\end{align}
\\ where
\begin{align}
\label{eq:effpotential}
V(x, D) = -D\, \ln \left[\frac{2 D}{G}\,
\sinh\left(\frac{G w(x)}{D} \right)
\, \right]\, ,
\end{align}
and the prime refers to the derivative with respect to $x$.
This equation describes the motion of a Brownian particle in a bistable
potential of entropic nature. Remarkably, the effective potential
does not only depend on the energetic contribution of the force $G$, but also on the
temperature and the geometry of the structure in a non-trivial
way. Notably, for the vanishing width at the two opposite corners of the geometry in Fig. (\ref{fig:well}) this entropic potential approaches infinity, 
thus intrinsically accounting for a natural reflecting boundary.
It is important to emphasize that this bistable potential was not present
in the 2D Langevin dynamics, but arises
due to the entropic restrictions associated to the confinement.
In general, after the coarse-graining the diffusion coefficient will depend on
the coordinate $x$ as well, but since in our case $|w^{\prime}(x)| \ll 1$,
this correction can be safely neglected, cf. Ref.  
\cite{Reguera_PRE, Burada_PRE, Zwanzig, Percus, Berezhkovskii2007}.

It is interesting to analyze the two limiting situations that can be obtained
upon varying the value of the ratio between the energy associated to the transversal
force and the thermal energy.
For the energy-dominated case,
i.e. $G w(x)/D\gg 1$,
Eq.~\eqref{eq:effpotential} yields
$V(x) = -G w(x)$ (neglecting irrelevant constants).
In the opposite limit, i.e. for $G w(x)/D \ll 1$,
the corresponding entropic potential function reads
$V(x,D) = -D\,\ln[2\, w(x)]$.

{\it Two-state approximation.-}\hspace*{0.05cm}
It is instructive to analyze the occurrence of stochastic resonance in the context
of the two-state approximation. For a potential $V(x)$ with barrier height $\Delta V$
the escape rate of an overdamped Brownian particle from one well
to the other in the presence of thermal
noise, and in the absence of a force, is given by the overdamped
Kramers rate \cite{McNamara,Jung91,hanggi},
reading in dimensionless units,
\begin{align}
\label{eq:krate}
r_\mathrm{K}(D) = \frac{\sqrt{V^{\prime \prime}(x_\mathrm{min})|V^{\prime \prime}
(x_\mathrm{max})|}}{2\pi}\,
\exp\left(\frac{-\Delta V}{D}\right) \, ,
\end{align}
where $V^{\prime \prime}$ is the second derivative of the
effective potential function,
and with $x_\mathrm{max}$ and $x_\mathrm{min}$ indicating the position
of the maximum and minimum of the
potential, respectively.

For the potential given by Eq.~\eqref{eq:effpotential} and the
shape defined by Eq.~\eqref{eq:widthfunctions},
the corresponding Kramers rate for transitions from one
basin to the other reads, in dimensionless units,
\begin{align}
  \label{eq:kr-both}
  r_\mathrm{K}(D) = \frac{G\,\epsilon}{\pi} \,
\frac{\sqrt{2\,\sinh\left(G b/D \right)\,
\sinh\left[G( b + 2 \epsilon)/D \right]}}
{{\sinh}^{2}\left[G( b + 2 \epsilon)/D\right]}\, .
\end{align}

{\it Spectral amplification.-}\hspace*{0.05cm}
The occurrence of Stochastic Resonance can be detected in the spectral
amplification $\eta$ \cite {Jung91}. It is defined by 
the ratio of the power stored in the response of the system at frequency
$\Omega$ and the power of the driving signal, and which for the periodically driven
two-state model, cf. Ref.~\cite{gammaitoni}, is given in dimensionless units as
\begin{align}
  \label{eq:amplification}
  \eta = \frac{1}{D^2}\, \frac{4\, {r^{2}_\mathrm{K}}(D)}{4\, {r^{2}_\mathrm{K}}(D) +
    \Omega^{2}}\, . 
\end{align}

Next we demonstrate the occurrence of the resonance in the
spectral amplification that signals the phenomenon of 
ESR. We demonstrate  that ESR  is neither a peculiarity of the
two-state approximation nor
of the equilibration assumption used to derive the effective
1D-Fokker-Planck equation.

In order to study the appearance of Stochastic Resonance we analyzed
the response of the system to the applied sinusoidal signal
in terms of the spectral amplification $\eta$.
In the presence of an oscillating force $F(t)$ in the $x-$direction there is
an additional contribution to the effective potential in
Eq.~\eqref{eq:fj} and the 1D kinetic equation in dimensionless units reads
\begin{align}
  \label{eq:fj-full}
  \frac{\partial P(x,t)}{\partial t} =
  \frac{\partial}{\partial x}\left\{D \frac{\partial P}{\partial x}  +
  \textbf{(} V^{\prime}(x,D) - F(t) \textbf{)}\,P \right\} \, .
\end{align}

The numerical integration of the 1D kinetic equation
\eqref{eq:fj-full} was done
by spatial discretization, using a Chebyshev collocation method, and
employing the method of lines to reduce the kinetic equation to a
system of ordinary differential equations, which was solved using a
backward differentiation formula method. This results in the
time-dependent probability distribution $P(x,t)$ and
the time-dependent mean value, defined as
\begin{align}
  \label{eq:meanx}
  \langle x(t) \rangle = \int x \, P(x,t) \mathrm{d} x \, .
\end{align}
In the long-time limit this mean value approaches the periodicity of driving \cite{Jung91}
with angular frequency $\Omega$. After a Fourier-expansion of $\langle x(t) \rangle $ one
finds the amplitude $M_1$ of the first harmonic of the output signal.
Hence, the spectral amplification $\eta$ for the fundamental
oscillation reads:
\begin{align}
\label{eq:samplification}
\eta = \left[\frac{M_1}{F_0} \right]^2\, .
\end{align}
\begin{figure}[t]
  \centering
  \includegraphics{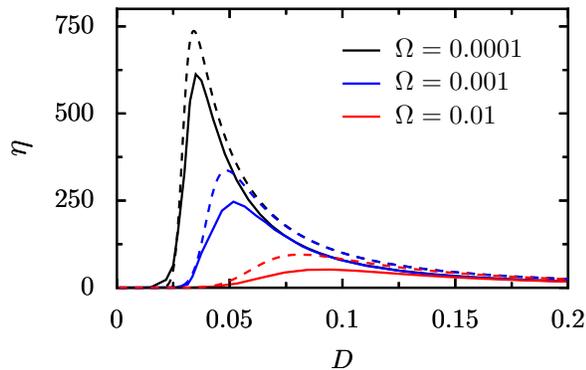}
  \caption{(color online) In dimensionless units,
    the dependence of the spectral amplification
    $\eta$ on noise level  $D$ for
    different driving frequencies,
    for the transversal force
    $G = 1.0$, the driving amplitude $F_{0}=10^{-4}$,
    and for the width function $w(x)=-\epsilon x^{4}+2 \epsilon x^{2}
    + 0.01 $ with
    the aspect ratio $\epsilon = 1/4$.
    The solid lines correspond to the 1D modelling,
    cf. Eq.\eqref{eq:fj-full} and Eq.\eqref{eq:samplification}, whereas
    the dashed lines correspond to the two-state approximation,
    cf. Eq.\eqref{eq:amplification}.}
  \label{fig:two-state}
\end{figure}

The comparison of the 1D modeling and the two-state
approximation in terms of the spectral amplification $\eta$,
cf. Eqs.~\eqref{eq:amplification} and \eqref{eq:samplification},
demonstrates the capability of the two-state approximation for small
driving frequencies and amplitudes, cf. Fig.~\ref{fig:two-state}.

{\it 1D modelling vs. precise numerics (2D).-}\hspace*{0.05cm}
In order to check the accuracy of the description
we compared the results obtained by the 1D modelling with
the results of Brownian dynamic simulations,
performed by integration of the overdamped Langevin
equation \eqref{eq:langevin}, for a 2D structure
(see Fig.~\ref{fig:well}) whose shape is described by
Eq.~\eqref{eq:widthfunctions}.
In our case we have used the aspect ratio $\epsilon = 1/4$ and
the bottleneck width $b = 0.02$.
The simulations were carried out by the
use of the standard stochastic Euler-algorithm.
\begin{figure}[!thb]
  \centering
  \includegraphics{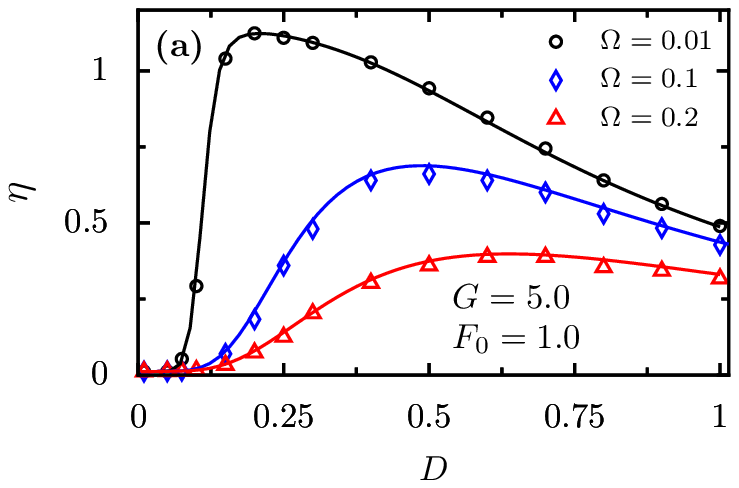}
  \includegraphics{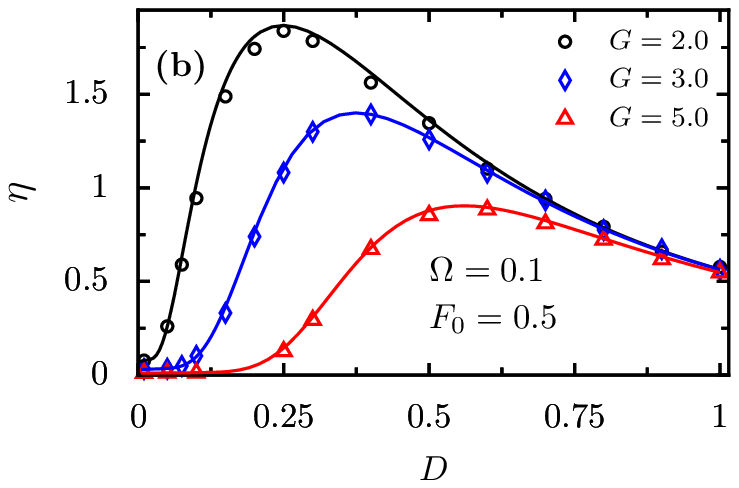}
  \includegraphics{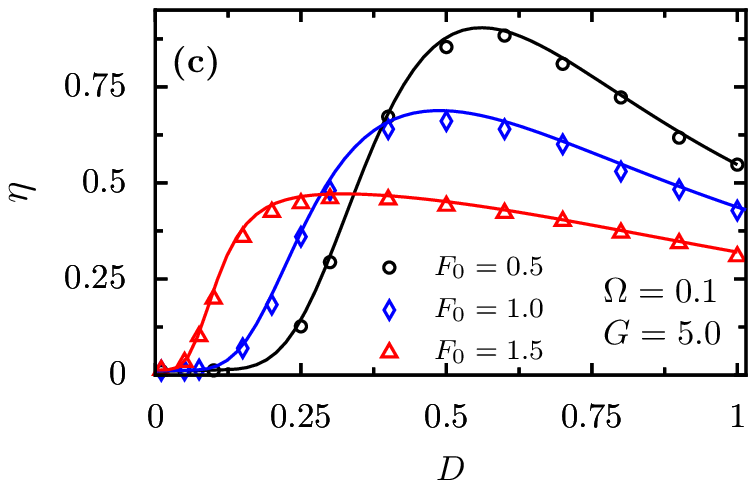}
  \caption{(color online) In dimensionless units,
    the dependence of the spectral amplification
    $\eta$ on noise level $D$ for various values of the
    quantities, the driving amplitude $F_0$, driving frequency $\Omega$, and
    the transversal force $G$.
    The symbols correspond to the results of the Langevin simulations
    for the two-dimensional structure with the shape defined by
    the dimensionless function
    $w(x) = -\epsilon x^4 + 2\,\epsilon x^2 + 0.01$
    with the aspect ratio $\epsilon = 1/4$,
    whereas the lines are the results of the numerical integration of
    the 1D kinetic equation \eqref{eq:fj-full}.}
  \label{fig:resonance}
\end{figure}

Fig.~\ref{fig:resonance} depicts the dependence of the spectral
amplification $\eta$ on the noise strength for different values of
the driving frequency, the driving
amplitude and the value of $G$. It is important to point out that
the results obtained from the 1D modelling using the 1D-approximation (lines)
are in excellent agreement with the numerical simulations
of the full system (2D) (symbols) within a small relative error.
This agreement is due to the fact that the considered potential function is a smooth
function ($|w^{\prime}(x)| \ll 1$), and in this situation our employed 1D-approximation
is expected to become very accurate \cite{Zwanzig, Reguera_PRE,Burada_PRE}.

Fig.~\ref{fig:resonance}a shows the dependence of the spectral
amplification $\eta$ on the noise strength $D$
for various driving frequencies at a fixed
transversal force and forcing amplitude $F_0$.
Here, we observe an increase in the spectral
amplification which gives rise to the finding of the effect of
Stochastic Resonance in the presence of entropic barriers.
As observed for  the usual "energy-dominated" SR
\cite{gammaitoni} the resonance peak is more pronounced as the applied angular
frequency $\Omega$ of the input signal decreases.
Similarly, Fig.~\ref{fig:resonance}b depicts how  ESR
depends on the strength of the transversal force $G$. Interestingly, the maximal amplification increases
upon decreasing the strength $G$ of the transversal force.
However, the presence of the transversal force $G$ is crucial for
observing a non-monotonic behavior of the spectral
amplification with increasing noise level $D$.
In the limit of  $G\rightarrow 0$, the spectral amplification
increases monotonically
with decreasing noise level and tends asymptotically
to $1/{\Omega}^2$ at small driving amplitudes $F_0$. Finally, 
Fig.~\ref{fig:resonance}c displays the dependence of the spectral
amplification $\eta$ on the noise strength $D$
for various amplitudes of the driving force $F_0$
at a fixed value of the transversal force and driving frequency.
Both, the amplification of the signal
and the optimal value of the noise at which it occurs seemingly 
increase as the driving amplitude
decreases.

This novel ESR effect is characterized by the appearance of a maximum in the
spectral amplification as a function of the noise strength $D$, just as in conventional energy-dominated SR \cite{gammaitoni}.
However, in biological and practical systems the temperature which controls
the strength of the thermal noise is not a readily variable parameter.
Our results suggest that, for a given temperature, a proper choice of the externally
controlled parameters (i.e. the nature of the driving force, i.e., its amplitude and driving frequency and
the strength of the transversal force) might bring the system into an optimal
regime where confinement and  noise mutually interplay to boost noise-assisted transport inside
a corrugated structure.

In conclusion, we have elucidated a new mechanism leading to the appearance of noise-induced
resonant effects when a Brownian particle moves in a confined medium in the
presence of  periodic driving. The constrained motion impedes the
access of the particle to certain regions of space and can be
described in terms of a bistable entropic potential. The activated dynamics of
the particle in this effective potential then results in a cooperative effect between noise
and external modulation,  yielding  an Entropic Stochastic Resonance. The
effect detected is genuine for small-scale systems in which shape and fluctuations
are unavoidable factors ruling their evolution. 
The advantageous possibilities of ESR 
on what concerns optimization and control may provide new perspectives in 
the understanding of systems at the scales of micrometers and
nanometers and open new avenues in their 
manipulation and control.

This work has been supported by the DFG via research
center, SFB-486, project A10, the
Volkswagen Foundation (project I/80424), the German Excellence
Initiative via the \textit {Nanosystems Initiative Munich} (NIM), and
by the DGCyT of the Spanish government through grant No. FIS2005-01299.

\end{document}